\documentstyle[twocolumn,aps,prl,graphicx,amssymb]{revtex}

\begin{document}

\draft

\title{Complex spatial organization in a simple model of 
resource allocation}

\author{Dami{\'a}n H. Zanette}
\address{Consejo Nacional de Investigaciones Cientificas y T\'ecnicas,
\\
Centro At\'omico Bariloche and Insituto Balseiro, 8400 Bariloche, 
R\'{\i}o Negro, Argentina}

\date{\today}

\maketitle

\begin{abstract}
A dynamical model for  the distribution of resources between competing
agents is   studied.    While global  competition leads to the
accumulation of all the resources by a single agent, local competition
allows for a wider resource distribution.  Multiplicative processes 
give  rise to almost-ordered spatial structures, thourgh the enhancement 
of random fluctuations.
\end{abstract}

\pacs{PACS numbers: 87.23.Ge, 87.23.Cc, 05.65.+b}

Though many real  systems in  the scopes of   biology and the   social
sciences    are well   described    by agent-based   models with  pair
interactions  \cite{Murray,Bak,wta}, qualitatively  similar to     the
physical  description  of  interacting  particles,  a  large class  of
biological and social processes are driven by interactions mediated by
an  external actor,  which  generally  bears  no resemblance with  the
individual agents. This  kind of processes---which, in connection with
physics, may  be  assimilated to   the evolution of   globally coupled
dynamical systems \cite{global}---is  typically found in systems where
agents compete for resources. A particularly relevant instance of this
situation is  given by competing biological  species. In this case, in
fact, the interaction is  rarely  given by struggling  events  between
individuals  of  different species,  but  rather by accessing  
simultaneously
to  the limited  resources   provided by the  environment.  The
performance of each  agent, i.e. of  each species, is here measured by
its ability to get resources, which depends both on the capabilities
of each individual and on global features such as the total population
of the species. The same scenario is found in some economical systems,
for instance, in  companies competing for financial resources, usually
administrated  by  banks, or  even   in  scientific research  projects
competing for funds from a government agency.  A key ingredient in the
dynamics of  these systems is  that the ability of each agent to get
resources at a given time depends, often strongly, on the resources
assigned to the   agent at previous  stages.  This can give rise  to a
multiplicative process that,   in  the absence of  buffer  mechanisms,
leads to resource accumulation by a single agent. In the following, we
consider  a simple dynamical   model that incorporates  these elements
and, in  particular, study the effects  of competition at local level.
We find that   local competition  softens   the process  of   resource
accumulation  and gives  rise,   through  the enhancement  of  spatial
fluctuations, to nontrivial structures.

Our  system consists   of an  ensemble  of $N$  agents,  each of  them
characterized by its productivity $\alpha_i$.  For future convenience,
we consider  the  generic situation    where all  productivities   are
different. At each time step $t$, each  agent is assigned an amount of
resources $r_i(t)$, which  is used  to produce  an amount of  products
$p_i(t)=  \alpha_i r_i(t)$.   At the  next  time step,   resources are
distributed among agents  in amounts proportional to their production,
namely,
\begin{equation} \label{model}
r_i(t+1) = \frac{p_i(t)}{\sum_j p_j(t)} R(t+1) =
\frac{\alpha_i r_i(t)}{\sum_j \alpha_j r_j(t)} R(t+1),
\end{equation}
where $R(t+1)$ are the total resources available at time $t+1$.  Model
(\ref{model}) can be fully solved for  arbitrary productivities
and  initial conditions $r_i(0)$. In   the first place,  we note  that
rescaling  $r_i(t)/R(t)   \to   r_i(t)$,  Eq.   (\ref{model})  becomes
independent of the total resources. We thus fix  $R(t)=1$ for all $t$,
so that  $\sum_j  r_j(t)=1$.  In   this  situation, the  solution   to
Eq. (\ref{model}) reads
\begin{equation} \label{solution}
r_i(t) = \frac{\alpha_i^t r_i(0)}{\sum_j \alpha_j^t r_j(0)}.
\end{equation}
For asymptotically large times, $r_i(t) \to 0$ for all $i$, except for
the agent   with the maximal    productivity, $\alpha_{\max} =  \max_i
\{\alpha_i \}$,  which receives all the  available  resources.  Due to
the    multiplicative effect   of   resource  allocation  according to
production, all resources are in the long run accumulated by the agent
with  the  maximal    productivity, giving     rise  to   a  sort   of
winner-takes-all state   \cite{wta}.    By  analogy  with   population
dynamics we say that the remaining  agents become extinct. In fact, in
connection  with biological populations,  this result is a realization
of  a  well-known   principle  of ecology, namely,    the principle of
competitive exclusion  \cite{theor,Murray}: in  a system of biological
species competing for the same resources, only one survives and the
others undergo extinction. At   moderately  large $t$  the   resources
assigned    to    each  agent        are    well   approximated     by
$r_i(t)=(\alpha_i/\alpha_{\max})^t r_i(0)/r_{\max}(0)$.  
Assuming that  the
productivities $\alpha_i$   are  drawn at random  from  a distribution
$P_\alpha (\alpha)$  and  that, for simplicity,  resources  are evenly
assigned   at   the   beginning,  $r_i(0)=N^{-1}$  for    all $i$, the
probability distribution for the  individual resources at a given time
is
\begin{equation} \label{P2}
P_r(r_i) = \frac{\alpha_{\max}}{r_it} r_i^{1/t} P_\alpha
\left( \alpha_{\max} r_i^{1/t} \right) .
\end{equation}
The dependence of this function on $r_i$ through the power $r_i^{1/t}$
is very weak for large $t$, so that on a wide interval of the variable
we find  $P_r(r_i) \propto r_i^{-1}$.  For long times, thus, resources
are distributed among agents following a power law with exponent $-1$.
Compare this  result    with  Pareto's  law  of wealth    distribution
\cite{Pareto,wta}.

Model (\ref{model}) admits several variations, that may be of interest
in connection with the description of some real systems. For instance,
the  extinction  of  all but  one agent   can be avoided  by modifying
slightly the   allocation  method. If,  at   each time  step,  a small
fraction $\rho$ of the total resources is evenly distributed among the
agents whereas the remaining  is  assigned according to production  as
above,  the individual resources are  always larger than $\rho/N$.  In
this  situation, $P_r(r_i)$ becomes  stationary for long times and, if
$\rho$ is small enough, the power-law dependence quoted above is found
again for $r_i>\rho/N$.

A second variation consists  in assuming that  the productivity can in
turn     depend on the  individual       resources, for instance,   as
$\alpha_i(r_i) =  \alpha_i^0 A(r_i)$.  
If
$A(r)$   increases    with $r$,   the    winner-takes-all   effect  is
enhanced. The opposite case, where $A(r)$ is a decreasing function, is
more interesting. Lower productivities for higher resources may be the
consequence of size effects, crowding,   or loss of efficiency due  to
lower competition. In this case,  the system  evolves to a  stationary
situation where the agents whose productivity weights $\alpha_i^0$ are
above a certain threshold $\alpha_{\min}^0$ receive nonzero resources,
whereas  the   agents with      $\alpha_i^0<\alpha_{\min}^0$    become
extinct.  In the stationary  situation,  all the surviving agents have
the same productivity $\alpha$. The third  instance, where $A(r)$ is a
nonmonotonous  function,   can  give   origin to multiple   stationary
nontrivial solutions.

In the following  we focus  the   attention on a variation  of   model
(\ref{model}) that introduces a spatial distribution of agents. Though
the  total resources are still  allocated among  all the ensemble, the
agents compete  at a local level only.  The set of agents that compete
with  a  given    agent     $i$  defines   its   neighborhood   ${\cal
N}_i$. Resources are assigned according to
\begin{equation} \label{local}
r_i(t+1) = \frac{1}{Z(t)}
\frac{\alpha_i r_i(t)}{\sum_{j  \in {\cal N}_i} \alpha_j 
r_j(t)}, 
\end{equation}
where 
\begin{equation} \label{Z}
Z(t) = \sum_i \frac{\alpha_i r_i(t)}{\sum_{j  \in {\cal N}_i} 
\alpha_j  r_j(t)}
\end{equation}
is a normalization factor that insures that $\sum_j r_j(t) =1$ for all
$t$. We   have   performed numerical  simulations   of  this system in
ensembles   of $N=10^3$   to  $10^4$ agents   distributed   on various
geometries. The productivities $\alpha_i$ were drawn  at random from a
uniform distribution in ($0,1$). Note  that, in Eqs. (\ref{model}) and
(\ref{local}), a homogeneous rescaling of all productivities leaves
the models invariant. Several distributions of initial conditions were
tested, but there are no  essential differences with the case where,
at   the    first     step,  resources      are  evenly     allocated,
$r_i(0)=N^{-1}$. Therefore, we  concentrate  our simulations on   this
simple case.

In the first place, we consider a one-dimensional array of agents with
periodic boundary conditions.  The   neighborhood of the  $i$th  agent
consists of its  two nearest neighbors. We find  that, for long times,
approximately half of  the  agents become  extinct  and resources  are
evenly allocated  among the  surviving agents.  Practically everywhere
along  the  array  agents are  ordered in   an alternating sequence of
extinct agents and  survivors. Occasionally,  one finds  two  neighbor
sites   where  both agents are  extinct,  but  two survivors are never
contiguous.  In other words, in the  neighborhood of a surviving agent
no  other survivor can be  found [cf. model (\ref{model})]. The
surprising feature of this asymptotic  distribution is that the strong
correlation between survival and  productivity found for Eq. (\ref{model})
is apparently   lost  in Eq. (\ref{local}).  In fact,   productivities are
distributed completely  at   random  along   the array,   whereas  the
resulting  distribution  of survivors  is  highly  ordered.   At first
glance,  no connection  can   be established,  for   instance, between
survivor  sites and local  maxima of  the productivity  or any spatial
pattern caused by fluctuations in the distribution of $\alpha_i$.

\begin{figure}
\centering
\resizebox{\columnwidth}{!}{\includegraphics{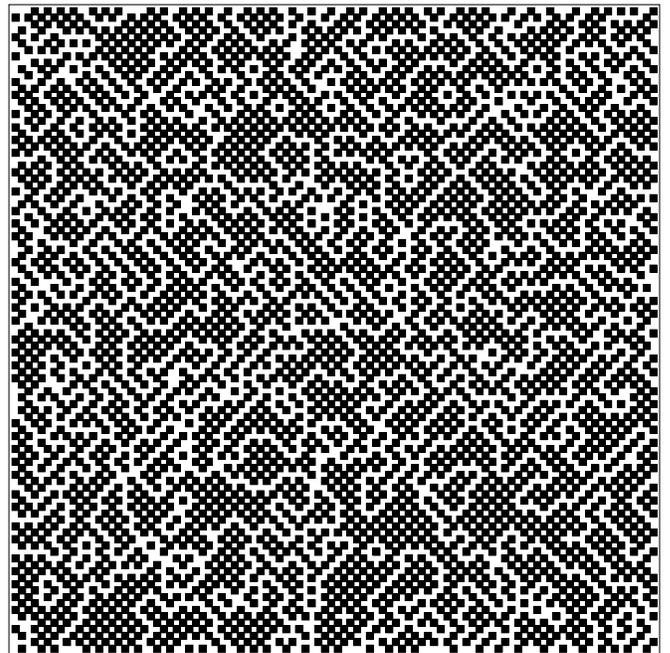}}
\caption{Asymptotic   state  on a  $100\times 100$-site   lattice with
nearest-neighbor competition and  periodic boundary  conditions.  Full
squares represent survivors.}
\label{2D}
\end{figure}

The  same feature is  found in  a  two-dimensional array with periodic
boundary conditions, where  the neighborhood of  each site consists of
the four  nearest neighbors. Figure  \ref{2D}  displays the asymptotic
state on  a $100 \times 100$-site  lattice ($N=10^4$),  where the full
squares represent survivor sites. Regular  domains where survivors and
extinct agents  alternate  in both dimensions, separated  by worm-like
boundaries formed by extinct agents, are apparent.  We have also 
verified  that different  definitions of the neighborhood of  a site, 
both in one and in  two-dimensional  lattices, produce similar
almost-periodic asymptotic structures. Again, resources are evenly 
distributed between the surviving agents and, in neighborhood of a 
survivor, no other survivor can be found. For instance, in  a
one-dimensional array  where  the neighborhood consists  of four sites
(nearest and next-to-nearest neighbors), the  resulting structure is a
periodic sequence of two extinct  agents and one survivor.  Occasional
defects, with larger zones of extinction, are also found.

In   order to trace   the  origin  of  the almost-periodic  structures
emerging at asymptotically long times we have inspected the successive
states of the  system  from  the earliest  evolution  stages, for  the
specific case of a one-dimensional array of  agents with two neighbors
per site  and periodic boundary   conditions. As above, productivities
are drawn at  random from a  uniform distribution and  $r_i(0)=N^{-1}$
for all $i$.  We find that even at the first step of the evolution the
distribution of resources already exhibits an almost-periodic sequence
of relatively  high  and  low  values,  in  spite   of the fact   that
productivities are  spatially uncorrelated.  Namely, for a substantial
fraction of agents,  we have $[r_{i+1}(1)-r_i(1)]  [r_i(1)-r_{i-1}(1)]
<0$.  This  observation  reveals that the    simple dynamics of  model
(\ref{local}) is able to  introduce strong spatial correlations  to an
initially uncorrelated state, even at the  level of a single evolution
step.

To quantify  this effect, we  take a  set of five  uncorrelated random
numbers  $\alpha_1,  \dots  ,\alpha_5$  drawn  from  the  productivity
distribution, and consider the combinations
\begin{equation} \label{comb}
\sigma_{i}=\frac{\alpha_i}{\alpha_{i-1}+\alpha_i+\alpha_{i+1}}
\end{equation}
for $i=2,3,4$. The quantities $\sigma_i$  are then proportional to the
resources received at $t=1$ by  three consecutive agents in the  array
[cf. Eq.  (\ref{local})]. Numerical realizations  of these quantities,
over  $10^6$ independent  choices of  $\alpha_1,\dots, \alpha_5$, show
that $(\sigma_4-\sigma_3)(\sigma_3-\sigma_2)<0$ with  probability $p_1
=  0.762$.  Consequently, slightly more than  76  \% of the agents are
expected to belong to a periodic sequence of  alternating high and low
resources at the first   time step. The  remaining  24 \% stands   for
defects in the periodic  structure.  The same  kind of analysis can be
performed for   successive steps in the  evolution.   It is found that
correlations  are  further enhanced by  the   dynamics.  At the second
evolution step,  for instance, the probability for  an agent to belong
to the  periodic structure grows to $p_2  = 0.782$.  For later stages,
we    find      $p_5=0.816$,  $p_{10}=0.839$,     $p_{20}=0.854$,  and
$p_{50}=0.864$. These probabilities   saturate  at $p_\infty   \approx
0.865$.

\begin{figure}
\centering
\resizebox{\columnwidth}{!}{\includegraphics{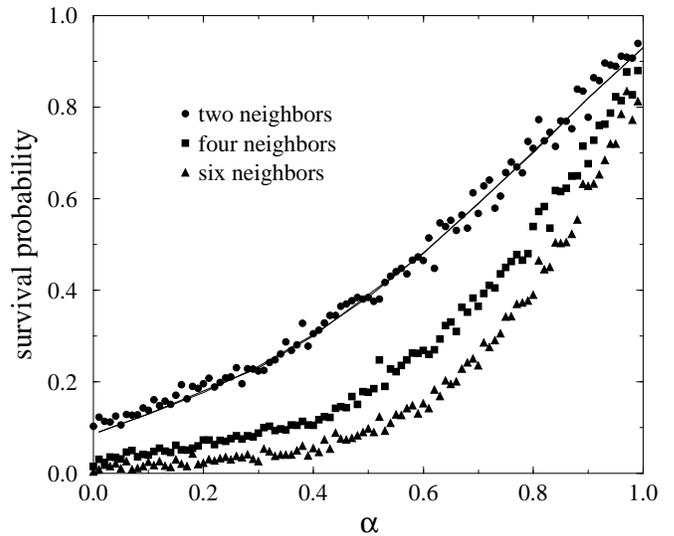}}
\caption{Survival  probability   as a   function  of the  productivity
$\alpha$ for   one dimensional arrays of   $1000$ agents and different
neighborhoods.   Dots correspond to   average measurements over  $100$
realizations.  The line shows the result of an independent calculation
method (see text), for the case of two neighbors.}
\label{prob1}
\end{figure}

While the  emergence of almost-periodic  structures seems  to indicate
that no correlations  subsist in model (\ref{model})  between survival
and productivity, the inspection of the evolution equations as well as
of Eq.  (\ref{comb})  suggests that some remaining  correlation should
however exist. In fact,  low productivities still imply few resources,
so that   the minima   in  the  periodic structure  should  preferably
coincide with sites  with small $\alpha_i$.   Moreover, defects in the
periodic structures---where extinct agents   are found in  consecutive
sites---should  also  correspond    to low-productivity  zones.   Such
remaining correlation  can be   characterized  by the probability   of
survival as  a  function of    the productivity, $p_s(\alpha)$.   This
probability is  defined as  the  fraction of agents with  productivity
$\alpha$    which  survive at   asymptotically   long  times. We  have
numerically measured  $p_s (\alpha)$, in   series of $100$ independent
realizations  of  the  productivity distribution    for ensembles with
$N=1000$, for various geometries and neighborhoods. Figure \ref{prob1}
shows   the  survival  probability as   a   function of   $\alpha$ for
one-dimensional arrays with different numbers of neighbors.  Here, the
correlation    between  survival   probability   and  productivity  is
apparent. Agents with larger productivities are more likely to survive
that those with small $\alpha$. Note, however, that for $\alpha=1$ the
survival probability  is  less than unity,  so that  even with the maximal
productivity there are chances of undergoing extinction. Conversely, for
$\alpha  \to 0$ (but   $\alpha\neq  0$), the survival probability   is
finite. As  the number  of neighbors grows    the probability is  more
concentrated towards larger productivities,  as expected. In the limit
where  the neighborhood extends to the  whole array the original model
(\ref{model}) is recovered,  and the survival probability must  vanish
except for $\alpha=1$.

The survival  probability  can be   calculated  independently of   the
numerical  realization     of the model,   using   expressions   as in
Eq. (\ref{comb}). For a fixed value of $\alpha_3$ and different random
choices  of $\alpha_1$,  $\alpha_2$,   $\alpha_4$ and $\alpha_5$,   an
estimate at the first evolution step of  the survival probability of
the  agent with  productivity $\alpha_3$  is given  by the fraction of
instances for which  $\sigma_3> \sigma_2, \sigma_4$. Generalizing this
procedure for successive time steps, the estimate can be improved by
considering   later    stages    in  the  evolution.     The  line  in
Fig. \ref{prob1} corresponds to this estimate at $t=50$. It shows an
excellent agreement with the numerical realizations.

\begin{figure}
\centering
\resizebox{\columnwidth}{!}{\includegraphics{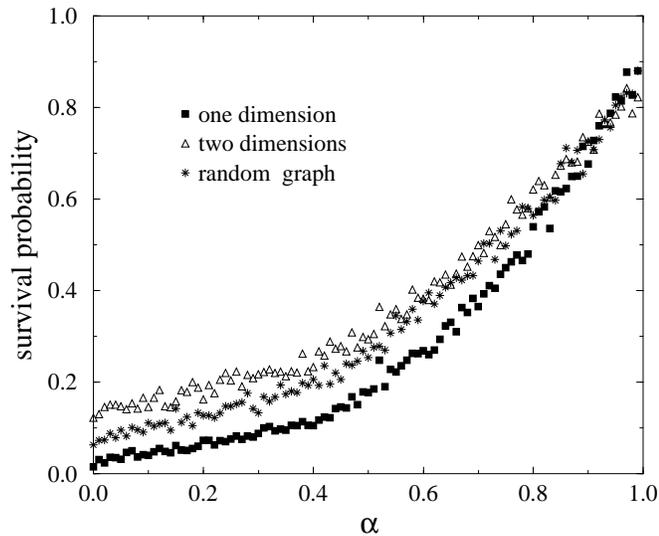}}
\caption{Survival probability  as  a   function  of the   productivity
$\alpha$ for different geometries with four neighbors per site.}
\label{prob2}
\end{figure}

To compare now   several  geometries, we  have  measured  the survival
probability  in  systems  where each  agent  has the  same   number of
neighbors---specifically, four---on different kinds of arrays, namely,
a  one-dimensional array (nearest  and  next-to-nearest neighbors),  a
two-dimensional  lattice (nearest neighbors)  and  a random graph with
four connections per site.  The   results, shown in fig.   \ref{prob2}
corresponds  to  averages  over $100$   realizations  in systems  with
$N=1000$ for one  dimension and random  graphs.  In  this latter case,
the   graph topology  is  chosen anew   at  each realization.  For two
dimensions, the  results  correspond to $10$   realizations on a  $100
\times 100$-array. Though the   general trend of  $p_s(\alpha)$  is
qualitatively the same for the  three cases, some systematic differences
are apparent.  For instance, the survival probability in one dimension
is    appreciably  lower  than  in  the    other  geometries for   low
productivities. In this range,  remarkably, the data for random graphs
lies  between those for  one  and two  dimensions. For $\alpha  >0.8$,
instead, the values of $p_s(\alpha)$ are hardly distinguishable within
our numerical precision.

We  may summarize  our main results   as  follows. First,  the strong,
deterministic  correlation between  survival and maximal  productivity
that characterizes the  model with global competition (\ref{model}) is
replaced, in the model with local competition (\ref{local}), by a much
weaker, probabilistic-like correlation.  Agents with high productivity
are more  likely to survive but, even   with the maximal productivity,
extinction cannot  be completely discarded.  On the other hand, agents
with very low productivity have  a chance of survival. This conclusion
should be relevant to the possible applications of the present models,
both to   economy  and to biology.    Second,  the loss   of the above
mentioned deterministic correlation is accompanied by the formation of
an  almost-regular spatial structure,   with a periodic alternation of
survivors  and extinct  agents.  This  structure is  explained by  the
emergence of strong spatial correlations out of the fully uncorrelated
distribution  of   productivities,  due to   the  very action   of the
evolution rules.    Nontrivial    correlations  in   the   conditional
probabilities  for certain combinations  of uncorrelated variables---a
rather  simple  but  counterintuitive phenomenon---have recently  been
pointed out for series of  random numbers \cite{Sornette}. The present
model illustrates the  occurrence of the same  kind of phenomenon in a
spatially extended dynamical system.

This work has been partially carried out at the Abdus Salam 
International Centre for Theoretical Physics (Trieste, Italy). The 
author thanks the  Centre for hospitality, and G. Abramson for
useful comments.

\end{document}